\documentclass[12pt]{iopart}
\usepackage{epsfig,color}
\usepackage{amssymb}
\usepackage{amsfonts}
\usepackage{amstext}
\usepackage{graphicx}

\newcommand{\U}[1]{{\omega}}

\def\mb{\mathbf}
\def\be{\begin{equation}}
\def\ee{\end{equation}}
\def\ba{\begin{eqnarray}}
\def\ea{\end{eqnarray}}

\newcommand{\ii}{{\mathrm{i}}}

\newcommand{\spp}{{s_{++}}}
\newcommand{\spm}{{s_{\pm}}}

\begin{document}

\title[Manifestation of impurity induced $s_{\pm}\Longrightarrow s_{++}$ transition...]{Manifestation of impurity induced $s_{\pm}\Longrightarrow s_{++}$ transition: multiband model for dynamical response functions }
\author{D.V.~Efremov}
\ead{efremov@fkf.mpg.de}
\address{Max-Planck-Institut f\"{u}r Festk\"{o}rperforschung, D-70569 Stuttgart,
Germany, IFW Dresden, Helmholtzstrasse 20, D-01069 Dresden, Germany}
\author{A.A.~Golubov}
\address{Faculty of Science and Technology and MESA+ Institute of Nanotechnology, University of Twente, 7500 AE Enschede,
The Netherlands}
\author{O.V.~Dolgov}
\address{Max-Planck-Institut f\"{u}r Festk\"{o}rperforschung, D-70569 Stuttgart, Germany}
\submitto{\NJP}
\date{\today}

\begin{abstract}
We investigate effects of disorder on the density of states, the single particle response function and  optical conductivity in  multiband superconductors with $s_\pm$ symmetry of the order parameter, where $s_{\pm} \to s_{++} $ transition may take place. In the vicinity of the transition the superconductive gapless regime is realized. It  manifests itself in anomalies in the above mentioned properties. As a result, intrinsically phase-insensitive experimental methods like ARPES, tunneling and terahertz spectroscopy may be used for revealing of information about the underlying order parameter symmetry.
\end{abstract}

\maketitle

\section{ Introduction}

The discovery of iron-based superconductors~\cite{kamihara} (FeSC) put forward
experimental and theoretical efforts to understand the reason for rather high critical temperatures and  symmetry of superconducting order parameters in these compounds. These studies
yielded a comprehensive experimental description of the electronic Fermi
surface structure, which includes multiple Fermi surface sheets in a good agreement with
density functional calculations~\cite{rev}.
The Fermi surface of the moderate doped FeSC is  given
by two small hole pockets around $\Gamma=(0,0)$ point and two
electron pockets around $M=(\pi,\pi)$ point in the folded zone. This band structure
suggests strong antiferromagnetic fluctuations, which may be a mechanism for
electron pairing. In this case, the natural order parameter for most of the FeSC is so called $s_{\pm}$ state, described by the nodeless order parameter with different signs for electron and hole-like pockets.

This model agrees well with the experimentally found for most of the moderately doped Fe-based superconductors nodeless character of the order parameter \cite{ARPES,popovich,hardy,charnukha}.
However a question, whether the order parameter changes its sign by changeover from  electron-like to hole-like  pockets, is still under discussion.
Moreover, relative robustness of the superconductors against nonmagnetic impurities
led to a suggestion that a more conventional two-band order parameter
a uniform sign change ($s_{++}$) is realized  in these systems~\cite{kontani}.

In our previous paper \cite{spm2spp} it was demonstrated  that not only superconductors with $s_{++}$  order parameter but also $s_{\pm}$ may be robust against nonmagnetic impurities.
Therefore the robustness against nonmagnetic impurities can not be considered as a strong argument against $s_\pm$ order parameter.
Moreover, it was shown, that
there are two types of $s_\pm$-superconductors with respect to disorder \cite{spm2spp}.
In  the first one $T_c $ goes down
as disorder is increased, until it vanishes at a critical value of
the scattering rate. This behavior is similar to the famous case of Abrikosov-Gor'kov magnetic impurities. It is widely discussed in the literature.
In the second type of $s_\pm$ superconductors $T_c$ tends to a finite value as
disorder is increased \cite{golubov97,golubov95}; at the same time the gap functions for the electron-like and hole-like Fermi surfaces acquire
the same  signs, i.e. the transition occurs from $s_\pm$ to $s_{++}$.

In the present paper we discuss how the disorder induced transition $\spm \to \spp$  can display itself in  single particle properties and optical conductivity.   It is shown that  disorder dependence of these characteristics  at the transition point is strongly
nonmonotonic function of the impurity scattering rate and can be easily seen in tunneling spectroscopy, ARPES and optics.
 Therefore systematic study of  disorder effects by means  formally phase insensitive techniques, mentioned above, may provide information about the sign of
the underlying order parameter in the clean limit.

The paper is organized as follows. In the section II we discuss the approximation used for the calculations. The section III is devoted to single particle
properties. We discuss how the transition $\spm \to \spp$ manifests itself in ARPES and tunneling spectroscopy. In the section III we calculate the two-particle response function and  discuss peculiarities  which can be seen in  optical conductivity in the vicinity of the transition point. The article is concluded with discussions.

\section{The formalism}
For the calculations we employ the standard approach of quasiclassical ($\xi$ - integrated) Green functions in Nambu and band space \cite{allen}:
\begin{equation}
\hat\mathbf{g}(\omega)=\left(
\begin{array}{cc}
\mathbf{g}_{a} & 0 \\
0 & \mathbf{g}_{b}%
\end{array}%
\right) ,  \label{eq.g}
\end{equation}
with band quasiclassical Green functions
\begin{equation}
\mathbf{g}_\alpha(\omega) = - i \pi N_\alpha \frac{\tilde\omega_\alpha \hat\tau_0 + \tilde\phi_\alpha \hat\tau_1 }{\sqrt{\tilde\omega_\alpha^2 - \tilde\phi^2_\alpha}},
\end{equation}
where the $\hat\tau_i$ denote Pauli matrices in Nambu space and $N_\alpha$ is the density of states on the Fermi level in the band $\alpha = a, b$  (for the sake of simplicity the two band model is considered).

The  function $\hat\mathbf{g}_\alpha$  is related to the
full Green function
\begin{equation}
\hat\mathbf{G}_\alpha(\mathbf{k},\omega_n)=\frac{ \tilde{\omega}_\alpha \hat\tau_0 + \xi_{\alpha} (\mb{k})  \hat\tau_3 + \tilde{\phi}_\alpha  \hat\tau_1 }{ \tilde\omega_\alpha^2-\xi_{\alpha}^2 (\mb{k}) - \tilde\phi_\alpha^2  }  \label{eq.Gf}
\end{equation}
by the standard procedure of $\xi$-integration
$\hat\mathbf{g}_\alpha(\omega) = N_\alpha \int d \xi_\alpha(\mathbf{k}) \hat G_\alpha(\mathbf{k},\omega)$.

The quasiclassical Green functions are obtained by numerical solution of the Eliashberg equations:
\begin{eqnarray}
&& \fl
\tilde \omega_\alpha(\omega)\!\! - \!\omega \! =   \!\!\!  \sum_{\beta = a,b} \left\{ \left.\int\limits_{-\infty}^{\infty} dz  K^{\omega}_{\alpha \beta} (z,\omega)
  Re  \frac{\tilde\omega_\beta(z) }{\sqrt{ \tilde\omega^2_\beta(z) - \tilde\phi^2_\beta(z) }}+
   i \Gamma_{\alpha \beta}(\omega)
  \frac{\tilde\omega_\beta(\omega) }{\sqrt{ \tilde\omega^2_\beta(\omega) - \tilde\phi^2_\beta(\omega) }}\right.
\label{eq.Elias.1}
  \right\},
  \\ &&
  \fl
\tilde \phi_\alpha(\omega) \!\! =  \sum_{\beta = a,b}  \left\{
\int\limits_{-\infty}^{\infty} dz  K^{\phi}_{\alpha \beta} (z,\omega)
 Re \frac{\tilde\phi_\beta(z) }{\sqrt{ \tilde\omega^2_\beta(z) - \tilde\phi^2_\beta(z) }}+i \Gamma_{\alpha \beta}(\omega)
\frac{\tilde\phi_\beta(\omega) }{\sqrt{ \tilde\omega^2_\beta(\omega) - \tilde\phi^2_\beta(\omega) }}
\label{eq.Elias.2}
  \right\},
\ea
where the kernels $K^{\tilde\phi, \tilde\omega}_{\alpha \beta}(z,\omega)$ have the standard from:
\be
K_{\alpha \beta}^{\tilde\phi, \tilde\omega} (z, \omega)   =
\int\limits_{-\infty}^\infty d \Omega \frac{\lambda_{\alpha \beta}^{\tilde\phi, \tilde\omega}  B(\Omega)}{2}  \times  \left[
\frac{\tanh \frac{z}{2T} + \coth \frac{\Omega}{2T}}{z+ \Omega-\omega - i \delta}
\right].
\ee
For simplicity we use the same normalized spectral function of electron-boson interaction  $B(\Omega)$ for all the channels, which is presented in inset Fig. 2. The maximum
of the spectra  is $\Omega_{sf} = 18$~meV \cite{charnukha}.   The matrix elements $\lambda^{\tilde\phi}_{\alpha\beta}$ are positive for attractive interactions and negative for repulsive ones. The symmetry of the order parameter in the clean case is determined solely by the off-diagonal matrix elements. The case  $sign \lambda^{\tilde\phi}_{a b} =sign \lambda^{\tilde\phi}_{b a} >0$ corresponds to $\spp$ superconductivity and $sign\lambda^{\tilde\phi}_{a b} =sign \lambda^{\tilde\phi}_{b a} <0$ to $\spm$.  The matrix elements
$\lambda^{\tilde\omega}_{\alpha\beta} $ have to be positive and  are chosen
$\lambda^{\tilde\omega}_{\alpha\beta} = |\lambda^{\tilde\phi}_{\alpha\beta}|$. For further calculations  we use the same matrix $\lambda^{\tilde\phi}_{aa}=3$, $\lambda^{\tilde\phi}_{bb}=0.5$, $\lambda^{\tilde\phi}_{ab}=-0.2$, $\lambda^{\tilde\phi}_{ba}=-0.1$ for $\spm$-case and $\lambda^{\tilde\phi}_{aa}=3$, $\lambda^{\tilde\phi}_{bb}=0.5$, $\lambda^{\tilde\phi}_{ab}=0.2$, $\lambda^{\tilde\phi}_{ba}=0.1$ for $\spp$-case. The correspondent ratio of the densities of states is $N_a/N_b = 0.5$ (see \cite{spm2spp}).

 The second terms in the right side of the Eqs.(\ref{eq.Elias.1}-\ref{eq.Elias.2}) reflect scattering on impurities.
In the general case $\Gamma_{\alpha \beta}(\omega)$ can be written in the following form:
\be
\Gamma_{\alpha \beta} (\omega ) = \gamma^N_{\alpha \beta} I(\omega),
\label{eq.gamma}
\ee
 where the $\gamma^N_{\alpha \beta}$ are inter- and intra-band impurity scattering rates in the normal state. The dynamical part $I(\omega) =1$ in Born approximation (see Appendix).  Beyond the Born approximation it reads:
\be
I(\omega) = \frac{1}{1 - 2 \zeta C_{ab}(\omega)},
\ee
 where  $C_{ab}(\omega)$ is the { \it coherence factor}:
 \be
C_{ab}(\omega) = 1 - \frac{\tilde\omega_{a}\tilde\omega_{b} - \tilde\phi_a\tilde\phi_b }{\sqrt{\tilde\omega_a^2 - \tilde\phi^2_a }\sqrt{\tilde\omega_b^2 - \tilde\phi^2_b }}.
\ee
Note that in the normal state $C_{ab}(\omega)  = 0 $ and $I(\omega) =1$.

The dimensionless constant $\zeta$ is related to the inter-band impurity scattering rate in the normal state $\gamma^N_{ab}$ as
\be
\gamma^N_{ab} = \frac{n_{imp}}{\pi N_a} \zeta.
\ee
The dependence of $\gamma^N_{\alpha, \beta}$ and $\zeta$ on the scattering potential is shown in the Appendix.

\section{Quasiparticle properties}
\subsection{DOS in superconductive state.}

Interband scattering is expected to modify the gap functions and the tunneling density of states (DOS) in
the superconducting state in a multiband superconductor. In the weak coupling regime the impurity effects have been
discussed in \cite{schopohl} within the Born limit and extended in \cite{dolgov2005} to the strong coupling case. In the following we will
calculate the gap functions, and the superconducting DOS by solving the nonlinear Eliashberg equations in the $s_{\pm}$ and $s_{++}$ superconductors
for  various values of the interband nonmagnetic scattering rate, going beyond the Born approximation.

Total DOS in superconducting state is given by the following expression
\begin{equation}
N(\omega)=\sum_{\alpha}N_{\alpha}(0)\mathop{\rm Re}\frac{\omega}{\sqrt
{\omega^{2}-\Delta_{\alpha}^{2}(\omega+i\delta)}}, \label{eq:dos}%
\end{equation}
where we have introduced the\textit{ complex order parameter}:
\be\Delta_{\alpha
}(\omega+i\delta)\equiv\omega\tilde{\phi}_{\alpha}(\omega+i\delta
)/\tilde{\omega}_{\alpha}(\omega+i\delta)=$ \textit{Re}$\Delta_{\alpha}%
(\omega)+i$ \textit{Im}$\Delta_{\alpha}(\omega).\ee
The solution
for $\Delta_{\alpha}(\omega)$ allows the calculation of the current-voltage
characteristic $I(V)$ and \textit{tunnelling conductance} $G_{NS}%
(V)=dI_{NS}/dV$ in the superconducting state of the $NIS$ tunneling junction.

In contrast to a single band case, where DOS does not depend on
\textit{nonmagnetic} impurities in the multiband case $\Delta_{\alpha}%
(\omega+i\delta)$ and DOS are strongly dependent on interband impurity scattering.

Figure 1 shows the calculated gap functions $\Delta_\alpha(\omega)$ for the bands $a$ and $b$ for different
interband impurity scattering rates.
One sees in both  $s_{\pm}$ and $s_{++}$ cases  strong non-monotonic  frequency dependence of the gap function with the maximums of the absolute values around $250 cm^{-1}$ for $a$-band $140 cm^{-1}$ for $b$-band, originated from the strong electron-boson coupling.
Furthermore,  the effects of impurity scattering are visible as
additional structure at low energies comparable to the interband scattering
rate $\Gamma_a$. The most spectacular effect is the impurity-induced sign change of $Re \Delta(\omega)$ at low energies in $b$-band
in the $s_{\pm}$ state. This $s_{\pm} \to s_{++}$ transition was predicted
in Matsubara representation in our earlier paper \cite{spm2spp}, where its consequences for non-vanishing $T_c$ vs disorder behavior were discussed.

\begin{figure}
[ptbh]
\begin{center}
\includegraphics[width=\linewidth] {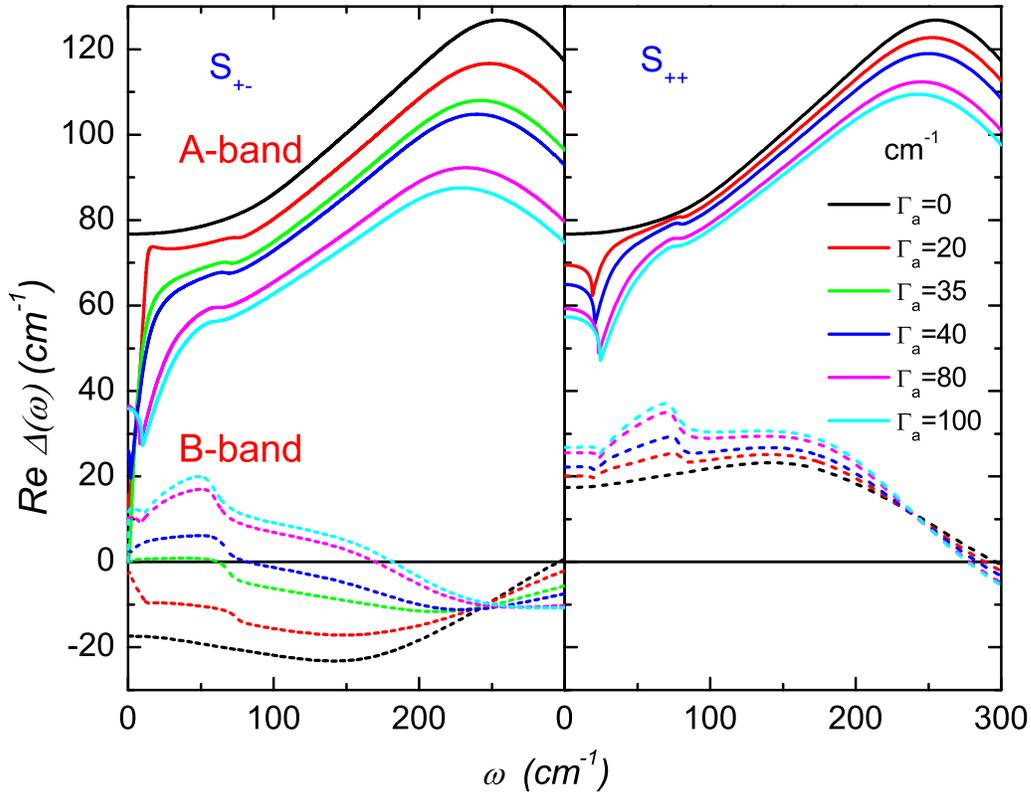}
\caption{Superconducting gap functions for bands $a$ and $b$ at various interband impurity scattering rates in $s_{\pm}$ and $s_{++}$ models.  The parameters are
$\zeta\approx0.2$,  $\gamma^N_{bb}=2\gamma^N_{aa} = \gamma^N_{ab}=2 \gamma^N_{ba} \approx 0.4  \Gamma_a $. They corresponds to scattering strength  $\sigma=0.5$.
 The relation between $\sigma$ and $\Gamma_{a}$ to the scattering potential is given in the Appendix.}
\label{fig1}
\end{center}
\end{figure}

\begin{figure}
[ptbh]
\begin{center}
\includegraphics[width=\linewidth] {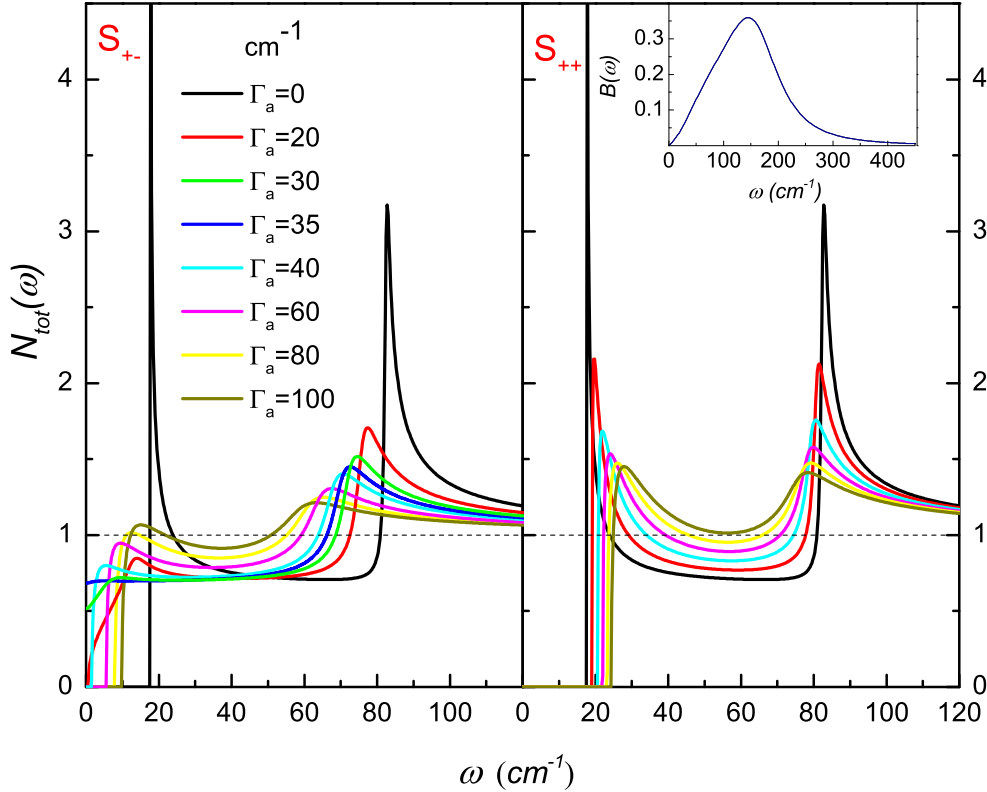}
\caption{Total density of states for various impurity scattering rates in $s_{\pm}$ and $s_{++}$ models at low temperature $T\ll T_{c0}$.
The parameters are the same as in Fig. 1. Inset shows the electron-boson interaction function $B(\Omega)$.}
\label{fig2}
\end{center}
\end{figure}

Figure 2 shows  comparison of DOS in $s_{\pm}$ and $s_{++}$ states for different magnitudes
of the interband scattering rate $\Gamma_a$ at low temperatures $T \ll T_{c0}$.
In the clean limit, one sees two different excitations gaps for the two bands. In accordance with earlier calculations for $s_{++}$ superconductors  \cite{dolgov2005}, the interband impurity scattering
mixes the pairs in the two bands, so that the states appear in the $a$-band
at the energy range of the $b$-band gap. These states are
gradually filled in with increasing scattering rate. At the
same time the minimal $b$-band gap in the DOS raises due to
increased mixing to the $a$-band with strong electron-boson coupling.
In the $s_{\pm}$ superconductor, the modifications of low-energy DOS with interband impurity scattering
is completely different. Due to sign change of $Re \Delta(\omega)$ in the $b$-band, gapless region exists
in a range of values of scattering parameter $\Gamma$ around 35 $cm^{-1}$, as clearly seen in the left panel in Fig.2.
Such gapless behavior manifests itself in optical properties of $s_{\pm}$ superconductors, as will be demonstrated below.

\subsection{ARPES and the self-energy}

Angle-resolved photo emission spectroscopy (ARPES) probes the photoemission
current $I(\mb{k},\omega)$, which can be calculated as:
\[
I(\mathbf{k},\U{3c9} )=\sum_\alpha|M_\alpha(\mathbf{k},\U{3c9} )|^{2}f(\U{3c9} )A_\alpha(\mathbf{k},%
\U{3c9} ).
\]
 Here  $M(%
\mathbf{k},\U{3c9} )$ is the dipole matrix element that depends on the initial and final electronic
states, incident photon energy and polarization, $f(\U{3c9} )$ is the
Fermi distribution function and
\be \fl A_\alpha(\mathbf{k}%
,\omega )=- \frac{1}{2 \pi} \Tr\left\{ Im \hat G_\alpha(\mathbf{k},\omega ) \hat\tau_0\right\} =
- \frac{1}{\pi} Im \frac{ \tilde{\omega}_\alpha(\omega)   }{ \tilde\omega_\alpha^2(\omega)-\xi_{\alpha}^2 (\mb{k}) - \tilde\phi_\alpha^2(\omega)  }
\label{eq.spectral.function}
\ee
is single particle response function.

In the weak coupling limit the contribution of the electron-boson interaction to self-energy $ \Sigma_{\alpha0}(\mb{k},\omega)$ \ (see first terms in l.h.s. of  Eqs. \ref{eq.Elias.1}, \ref{eq.Elias.2}) vanishes. It means, that in the model with isotropic self-energy
$\Sigma _{\alpha
0}^{e-b}(\omega )\rightarrow 0$, $\Sigma _{\alpha 1}^{e-b}(\omega
)\rightarrow \Delta _{\alpha }(\omega )$. Then the single particle spectral function takes the form:
\[
A_{\alpha }(\mb{k},\omega)=\frac{1}{\pi } Im\frac{\omega \left[ 1+ i\sum\limits_{\beta
=a,b} \Gamma _{\alpha \beta }/\sqrt{\omega ^{2}-\Delta _{\beta
}^{2}(\omega )}\right] }{D}
\]%
with
$$
D=\xi _{\alpha }^{2}(\mb{k})+ \left( \sum\limits_{\beta
=a,b}\Gamma _{\alpha \beta }\right) ^{2}-\omega ^{2}+\Delta _{\alpha
}^{2}(\omega )-2i\sum\limits_{\beta =a,b} \Gamma _{\alpha \beta } \frac{
\left[ \omega ^{2}+\Delta _{\alpha }(\omega )\Delta _{\beta }(\omega )\right]}
{\sqrt{\omega ^{2}-\Delta _{\beta }^{2}(\omega )}}.
$$
In the gaped regime $A_\alpha(\mb{k}, \omega)$ vanishes below $\Delta _{\beta }$ but in the
gapless one for $b$-band is the same as in the normal state:
\[
A_{b }(\mb{k},\omega)=\frac{1}{\pi }Im\frac{\omega\left[ 1 +i\sum\limits_{\beta
=a,b}\Gamma _{b \beta }/|\omega|\right]}{\xi _{b }^{2}(\mb{k})-\omega^2 \left(1+i  \sum\limits_{\beta
=a,b}\Gamma _{b \beta }/|\omega| \right) ^{2}}.
\]%

\begin{figure}
[ptbh]
\begin{center}
\includegraphics[width=0.9\linewidth] {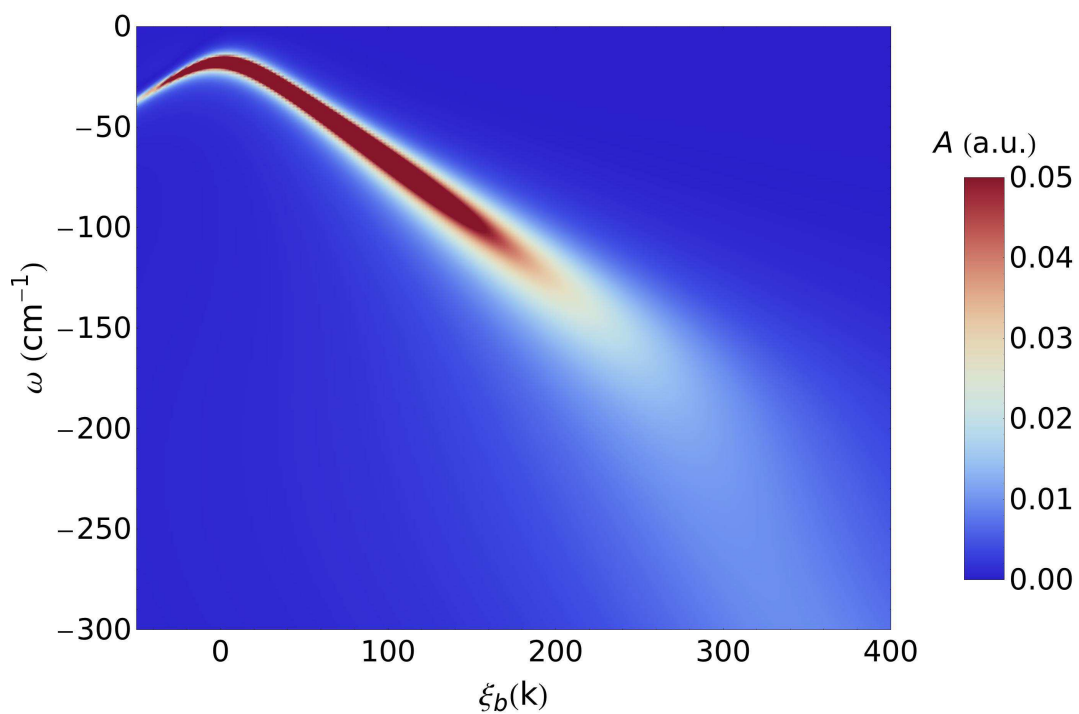}
\caption{The quasi-particle spectral function $A_{b}(\mathbf{k},\omega)$
\ for $b$-band with a small gap in the clean limit. The parameters are the same as in Fig. 1.}
\label{fig3a}
\end{center}
\end{figure}

The quasi-particle spectral function $A_{b}(\mathbf{k},\omega )$ given by Eq.(\ref{eq.spectral.function}) for $b$-band is shown in Fig.3. In this case the behavior of $A_{b }(%
\mathbf{k},\omega )$ at small $\omega $ and $\xi $ reflects the existence of
well-defined energy gap. In contrast to that, the function $A_{b}(\mathbf{k}%
,\omega )$ in the regime of $s_{\pm }\rightarrow s_{++}$ transition shows no
gap, as seen from Fig.4. With further increase of scattering rate $%
\Gamma _{a}$, when $s_{++}$ state is realized, in the $b$-band energy gap appears again.
Therefore, ARPES measurements at various impurity concentrations may provide
useful tool to distinguish underlying pairing symmetry of superconducting
state in pnictides.

\begin{figure}
[ptbh]
\begin{center}
\includegraphics[width=0.9\linewidth] {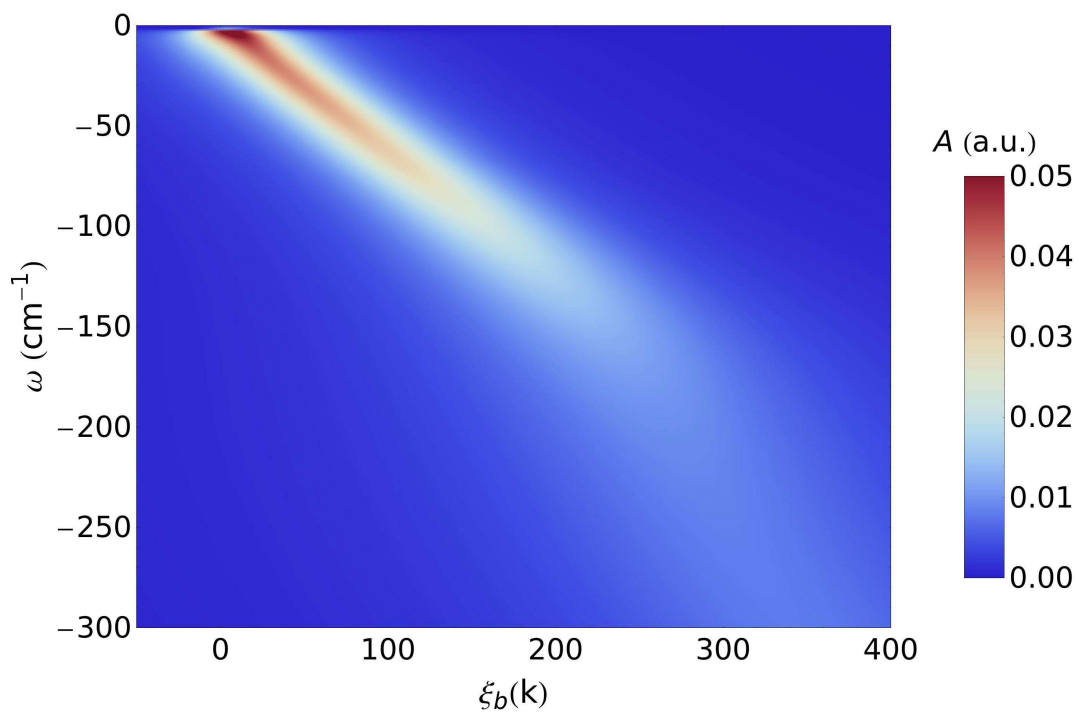}
\caption{The quasi-particle spectral function $A_{b}(\mathbf{k},\omega)$
for $b$-band with a small gap in the gapless regime ($\Gamma_a=40 cm^{-1}$). The parameters are the same as in Fig. 1.}
\label{fig4a}
\end{center}
\end{figure}

\section{Optical conductivity}

    The optical conductivity in the London (local, $\mathbf{q}\equiv0$) limit  in {\it a-b} plane
is given by
\begin{equation}
\sigma(\omega)=\sum_{\alpha}\omega_{pl,\alpha}^{2}\Pi_{\alpha}(\omega)/4\pi
i\omega, \label{sig}%
\end{equation}
where $\Pi_{\alpha}(\omega)$ is an analytical continuation to the real
frequency axis of the polarization operator (see, e.g. Refs. \cite{Nam}%
,\cite{LRZ},\cite{DGS},\cite{AC},\cite{mars1})
\[
\Pi_{\alpha}(\omega)=\left\{  i\pi T\sum_{n}\Pi_{\alpha}(\omega'_{n},\nu
_{m})\right\}  _{i\nu_{m}\Longrightarrow\omega+i0^{+}},
\]
$\alpha=a,b$ is the band index. %

\begin{eqnarray} \fl
\Pi _{\alpha}(\omega )   =
\int d\omega ^{\prime }\left\{ \frac{\tanh \left( \frac{%
\omega _{-}}{2T}\right) }{D^{R}}
\right. \left[ 1-\frac{\tilde{\omega}%
_{-}^{R}\tilde{\omega}_{+}^{R}  + \tilde\phi _{-}^{R}\tilde\phi _{+}^{R}}
{Q_{-}^R Q_{+}^R }
\right]
  \label{real}
-\frac{\tanh \left( \frac{\omega _{+}}{2T}\right) }{D^{A}}\left[ 1-\frac{%
\tilde{\omega}_{-}^{A}\tilde{\omega}_{+}^{A}+ \tilde\phi _{-}^{A}\tilde\phi
_{+}^{A}}
{Q_{-}^A Q_{+}^A }
\right]
\nonumber \\
-\frac{\tanh \left( \frac{\omega _{+}}{2T}\right) -\tanh \left( \frac{%
\omega _{-}}{2T}\right) }{D^{a}}\left. \left[ 1-\frac{\tilde{\omega}%
_{-}^{A}\tilde{\omega}_{+}^{R}+ \tilde\phi _{-}^{A}\tilde\phi _{+}^{R}}
{Q_{-}^A Q_{+}^R }\right] \right\}
,
\end{eqnarray}%

where
$$
Q^{R,A}_{\pm} = \sqrt{%
(\tilde{\omega}_{\pm}^{R,A})^{2}\mathbf{-(}\tilde\phi _{\pm}^{R,A})^{2}},
$$

\[
D^{R,A}=\sqrt{(\tilde{\omega}_{+}^{R,A})^{2}\mathbf{-(}\tilde{\phi} _{+}^{R,A})^{2}}+%
\sqrt{(\tilde{\omega}_{\_}^{R,A})^{2}\mathbf{-(}\tilde{\phi} _{-}^{R,A})^{2}},
\]%
and
\[
D^{a}=\sqrt{(\tilde{\omega}_{+}^{R})^{2}\mathbf{-(}\tilde{\phi} _{+}^{R})^{2}}-\sqrt{%
(\tilde{\omega}_{\_}^{A})^{2}\mathbf{-(}\tilde{\phi} _{-}^{A})^{2}},
\]%
$\omega _{\pm }=\omega ^{\prime }\pm \omega /2$, and the index $R(A)$
corresponds to the retarded (advanced) brunch of the complex function $%
F^{R(A)}=Re F\pm iImF$ ( the band index $\alpha$ is omitted).

In the normal state the conductivity is
$$
\sigma_\alpha^N(\omega)  = \frac{\omega_{pl}^2}{8 \ii \pi \omega}   \int_{-\infty}^\infty dz
 \frac{ \tanh((z+\omega)/2T) -  \tanh(z/2T) }{\tilde\omega_\alpha(z+\omega)-\tilde\omega_\alpha(z)  }.
$$
If the dominant contribution to the quasiparticle damping comes from the impurity scattering,  it reduces to the Drude formula:
$$
\sigma_a(\omega,T) = \frac{\omega_{pl}^2}{4 \pi} \frac{1}{\gamma^{opt}_a -i \omega  }
$$
with
$
\gamma^{opt}_a = \gamma_{ab}+ \gamma_{aa}.
$

In the Fig.~\ref{fig5} we demonstrate the impact of disorder on the optical conductivity $Re\sigma(\omega)$. In the clean limit one sees
$Re\sigma_\alpha(\omega) =0 $ for  $\omega<2\Delta_\alpha$.
With increase the impurity scattering rate the region $Re\sigma_b(\omega) =0$ for the band $b$ decreases and the peak above $2\Delta_b$ becomes sharper.
It is clearly seen that in the vicinity of transition  from the $s_\pm$ to $s_{++}$ state ($\Gamma_a \sim 35 cm^{-1}$), the conventional Drude-response characteristic in the weak for a normal metal-state is realized. The origin of this effect is gapless nature of superconductivity near the impurity-induced $s_\pm \to s_{++}$ transition.
With further increase of the impurity scattering rate, the optical conductivity recovers gapped-like behavior, but with smaller gap. It is strikingly different from the behavior of superconductors with $s_{++}$  order parameter (Fig. \ref{fig5}.b), where values of two gapes tends to merge at the limit of infinite impurity scattering rate.
This reentrant behavior of optical conductivity with concentration of nonmagnetic impurities may serve as unambiguous indication for the $s_{\pm}$ order parameter symmetry.

\begin{figure}
[ptbh]
\begin{center}
\includegraphics[
width=\linewidth
]{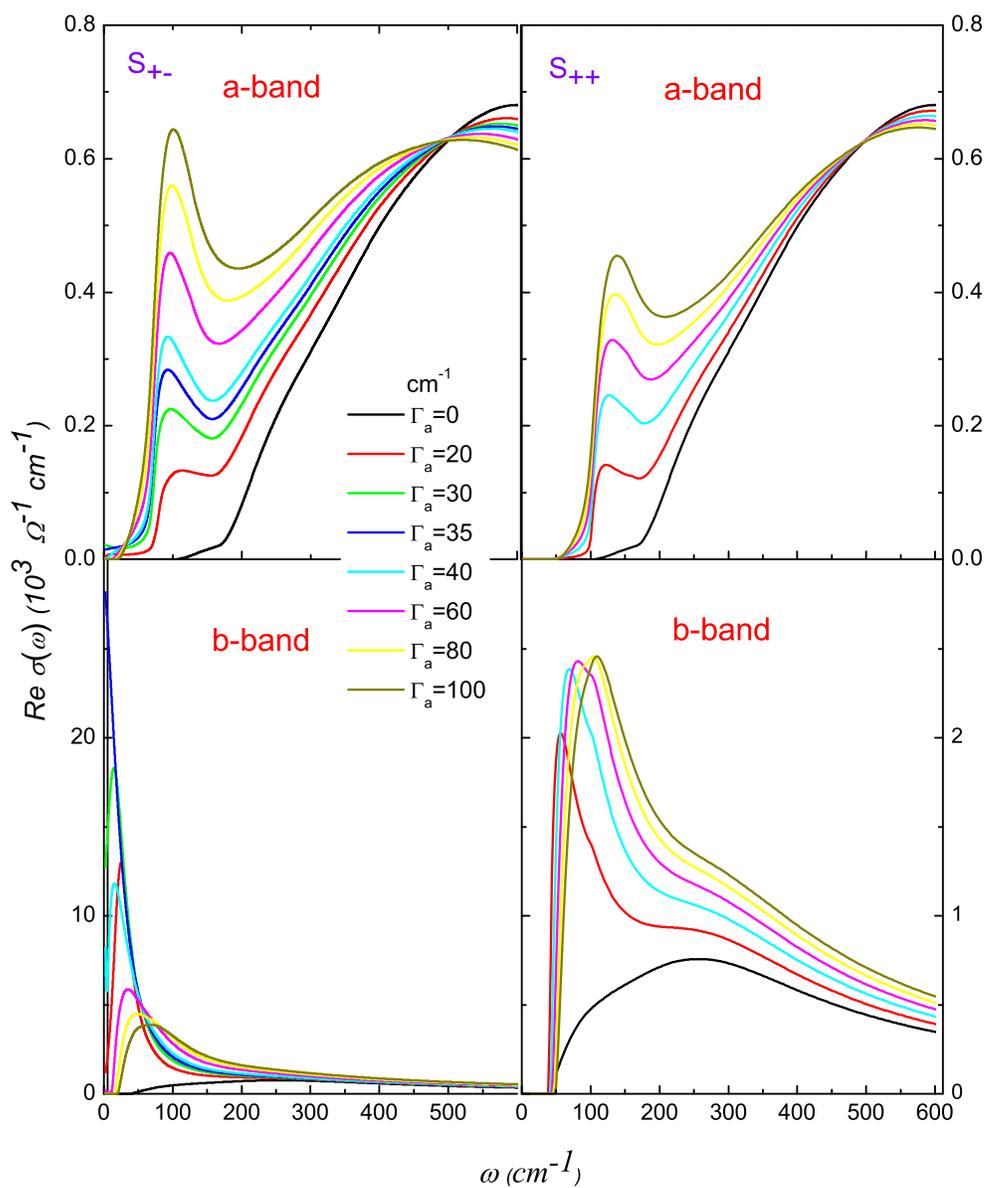}%
\caption{Real part of the optical conductivity $Re\sigma(\omega)$ for various values of $\Gamma_a$. The parameters are the same as in Fig. 1. }%
\label{fig5}%
\end{center}
\end{figure}

\begin{figure}
[ptbh]
\begin{center}
\includegraphics[
width=\columnwidth
]{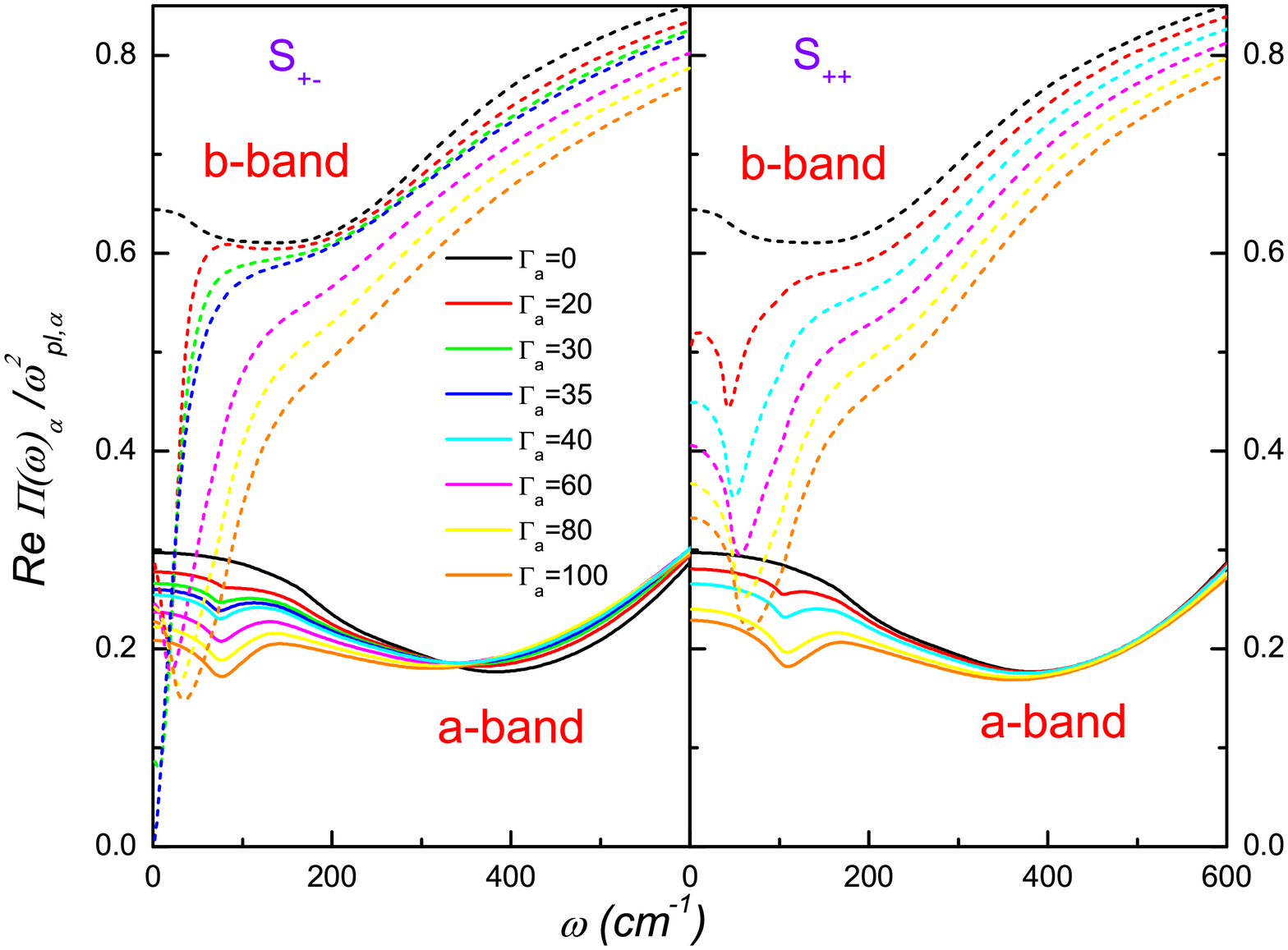}%
\caption{Real part of the polarization operator for various values of $\Gamma_a$. The parameters are the same as in Fig. 1.}%
\label{fig6}%
\end{center}
\end{figure}

Another important characteristic of the superconducting state is the real part of the electromagnetic kernel (polarization operator)
which is related to the imaginary part of optical conductivity $Im\sigma(\omega)$ (see Eq.\ref{sig}).
Fig.6 shows the frequency dependence of Re $\Pi(\omega)$ for $s_{\pm}$ and $s_{++}$ models
for various interband scattering rates. One can see that in the $s_{++}$ case dips at $\omega=2 \Delta_\alpha(\omega)$ occur for nonzero
scattering, in accordance with previous calculations done for single-band superconductors \cite{marsiglio}.
Further, an interesting peculiarity is seen in the response of the $b$-band in the $s_{\pm}$ state: the dip position is a nonmonotonic function of interband scattering rate and
the dip vanishes completely in the gapless regime corresponding to the $s_{\pm} \to s_{++}$ transition.

Magnetic field penetration depth   $\lambda_{L}(T)$ in the local (London)
limit in {\it a-b} plane is related to $Im\sigma(\omega)$ in zero frequency limit
\begin{equation}
1/\lambda_{L}^{ 2}(T)=\sum_{\alpha=a,b}\lim_{\omega \to 0}4\pi\omega
\mathop{\rm Im}\sigma_{\alpha}(\omega,\mathbf{q}=0,T)/c^{2},
\label{eq:pen-london}%
\end{equation}
where  $c$ is the velocity of light.
If we neglect strong-coupling effects (or, more generally, Fermi-liquid
effects) then for a clean uniform superconductor at $T=0$ we have the relation
$\lambda_{L}=c/\omega_{pl},$ where $\omega
_{pl, \alpha}=\sqrt{8\pi e^{2}\langle N_{\alpha}(0)v_{F\alpha}v_{F\alpha}\rangle}$ is
the plasma frequency in different bands.

Partial contributions to the magnetic field penetration depth can be written
as
\ba \fl
1/\lambda_{L}^{2}(T)=Re \sum_{\alpha=a,b}\frac{\omega_{pl,\alpha}^{2}}{c^{2}%
}  \int_{\omega_{g}(T)-0}^{\infty}\frac{d\omega \tanh\left(  \omega/2T\right)}%
{Z_{\alpha}(\omega,T)\sqrt{\omega^{2}-\Delta_{\alpha}^{2}(\omega,T)}}  \frac{\Delta
_{\alpha}^{2}(\omega,T)}{\left[  \Delta_{\alpha}^{2}(\omega,T)-\omega^{2}\right]  }%
    .
\ea
Here the points $\omega_{g}(T)$ are determined by the condition for the density of states in the band
$$
 Re N(\omega <\omega_{g}(T))=0.
$$
For superconductors with gap nodes as well as for $T>0$:  $\omega_{g}(T) \equiv 0$ (see, \cite{Maksimov}).

Peculiarities of the penetration depth in the crossover regime from $s_{\pm}$ to $s_{++}$ state
have been discussed earlier \cite{spm2spp}.

\section{Conclusions}

We have studied the effects of the impurity-induced  $s_{\pm} \to s_{++} $ transition in the density of states, the single particle response function and optical conductivity in a multiband superconductors with $s_\pm$ symmetry of the order parameter. It has been  shown that smaller gap vanishes in the vicinity of this transition, leading to gapless nature of photoemission and tunneling spectra. In optical response, the $s _{\pm} \to s_{++} $ transition leads to "restoring " of the \textquotedblleft
Drude\textquotedblright-like frequency dependence of $Re\sigma(\omega)$. We have also found interesting anomalies in the real part of polarization operator, with reentrance behavior of the dip-like structure at $\omega=2\Delta_\alpha(\omega)$ as a function of interband scattering rate. This effect leads to non-monotonic behavior in the magnetic field penetration depth as a function of the impurity concentration.

We want to stress that systematic study of the impact of disorder  on the single-particle response function and optical conductivity may give an information about underlying symmetry of the superconductive order parameter.

\ack The authors are grateful to P.~Hirschfeld, M.M.~Korshunov, A.~Charnukha,
A.V.~Boris, B.~Keimer for useful discussions. The
present work was partially supported by the DFG Priority Programme SPP1458
(DVE), Dutch FOM (AAG).

\appendix
\section*{Appendix}
\setcounter{section}{1}
 The impurity part of the self energy is obtained by analytic continuation of the correspondent self-energy on Matsubara frequencies
 $\hat\Sigma_{\alpha \beta}^{imp}(i\omega_n) = \Sigma_{\alpha \beta}^{imp(0)}(i\omega_n) \hat\tau_0 + \Sigma_{\alpha \beta}^{imp(3)}(i\omega_n) \hat\tau_3 $ derived in \cite{spm2spp}. The last is taken in the $T$ - matrix approximation:
\begin{equation}
\hat\mathbf{\Sigma}^{imp}(\omega_n) =n_{imp} \hat\mathbf{U} + \hat\mathbf{U} \hat%
\mathbf{g}(\omega_n) \hat\mathbf{\Sigma}^{imp}({\mathrm{i}}\omega_n),
\label{eq.tmatrix}
\end{equation}
where $\hat\mathbf{U} = \mathbf{U} \otimes \hat\tau_3$ and $n_{imp}$ is
the impurity concentration and the scattering potential:
$$
\mathbf{U} = \left(
\begin{array}{cc}
U_{aa} & U_{ba} \\
U_{ab} & U_{bb}
\end{array}
\right).
$$

The quasiclassical Nambu Green's functions on Matsubara frequencies are:
\begin{equation}
g_{0\alpha} = - \frac{{\mathrm{i}} \pi N_\alpha \tilde\omega_{\alpha n}} {%
\sqrt{\tilde\omega_{\alpha n}^2 + \tilde\phi_{\alpha n}^2}}, \;\;
g_{1\alpha} = - \frac{ \pi N_\alpha \tilde\phi_{\alpha n}} {\sqrt{%
\tilde\omega_{\alpha n}^2 + \tilde\phi_{\alpha n}^2}}.  \label{g}
\end{equation}
Solutions of Eqs.~(\ref{eq.tmatrix}) for $\Sigma_{aa}^{imp(0)}$ and
$\Sigma_{aa}^{imp(1)}$ are
\begin{eqnarray}
\Sigma_{aa}^{imp(0)}    = &n_{imp} \frac{   U_{aa}^{2}%
-\left(  \det \mathbf{U} \right)  ^{2}\left(  g_{0b}^{2} -g_{1b}^{2}\right)}{D(\omega_{n})}\, g_{0b}  + n_{imp}\frac{    U_{ab} U_{ba} }{D(\omega_{n})} \,g_{0b},\label{eq.SigmaGen0}\\
\Sigma_{aa}^{imp(1)}    = -&n_{imp} \frac{  U_{aa}^{2}%
-\left(  \det \mathbf{U}\right)  ^{2}\left(  g_{0b}^{2}-g_{1b}^{2}\right)  }{D(\omega_{n})} \,g_{1a}
-n_{imp}\frac{   U_{ab} U_{ba}  }{D(\omega_{n})} \,g_{1b}, \label{eq.SigmaGen1}%
\end{eqnarray}
where
\begin{eqnarray}
\fl
D(\omega_{n})=1-\left(  g_{0a}^{2}-g_{1a}^{2}\right)  U_{aa}^{2} -\left(  g_{0b}^{2}-g_{1b}^{2}\right)  U_{bb}^{2} \nonumber
+\left(  g_{0a}%
^{2}-g_{1a}^{2}\right)  \left(  g_{0b}^{2}-g_{1b}^{2}\right)
 \left(  \det
\mathbf{U}\right)  ^{2} \nonumber
\\  -2U_{ab}U_{ba}\left(
g_{0a}g_{0b}-g_{1a}g_{1b}\right) .
\end{eqnarray}
\begin{figure}
[ptb]
\begin{center}
\end{center}
\end{figure}

The analytical continuation on the real axis leads to the following expression:
$$
\hat\mathbf{\Sigma}^{imp (0)}(\omega) =  \sum_{\beta = a,b}
   i \Gamma_{\alpha \beta}(\omega)
  \frac{\tilde\omega_\beta(\omega) }{\sqrt{ \tilde\omega^2_\beta(\omega) - \tilde\phi^2_\beta(\omega) }},
$$
and
$$
\hat\mathbf{\Sigma}^{imp (1)}(\omega) =  \sum_{\beta = a,b}
   i \Gamma_{\alpha \beta}(\omega)
  \frac{\tilde\phi_\beta(\omega) }{\sqrt{ \tilde\omega^2_\beta(\omega) - \tilde\phi^2_\beta(\omega) }}
$$
with $\Gamma_{\alpha \beta}(\omega)$ given by Eq.(\ref{eq.gamma}).

The dimensionless constant $\zeta$ is convenient to express  in terms of dimensionless  scattering potentials $\bar{u}_{\alpha \beta} = \pi U_{\alpha \beta} N_\beta $ and $\bar{d} = \bar{u}_{aa} \bar{u}_{bb} - \bar{u}_{ab} \bar{u}_{ba}$. Then it has the following compact form:
\be
\zeta = \frac{
 \bar{u}_{ab}  \bar{u}_{ba}}{ (\bar{d}-1)^2 + (\bar{u}_{aa}+\bar{u}_{bb})^2 }.
 \label{eq.app.zeta}
\ee
Normal state impurity scattering rate reads:
\be
\gamma^N_{\alpha \alpha} = \frac{n_{imp}}{\pi N_\alpha }\frac{\bar{d}^2 + \bar{u}_{\alpha \alpha}^2}{ (\bar{d}-1)^2 + (\bar{u}_{aa}+\bar{u}_{bb})^2}
\label{eq.app.gamma.aa}
\ee
and for $\alpha \ne \beta$
\be
\gamma^N_{\alpha \beta} = \frac{n_{imp}}{\pi N_\alpha } \frac{
 \bar{u}_{ab}  \bar{u}_{ba}}{ (\bar{d}-1)^2 + (\bar{u}_{aa}+\bar{u}_{bb})^2 }.
\label{eq.app.gamma.ab}
\ee
The Born approximation corresponds to $\bar{u}_{\alpha \beta} \ll 1$. Then up to quadratic terms in $\bar{u}$ one gets:
$$
\Gamma_{aa}(\omega) \approx \gamma^N_{aa} \approx n_{imp} \pi N_a U_{aa}^2
$$
and
$$
\Gamma_{ab}(\omega) \approx \gamma^N_{ab} \approx n_{imp} \pi N_b U_{ab}^2.
$$

It is worth to introduce  the parameters 
 $\sigma = \bar{u}_{ab} \bar{u}_{ba}/(1 + \bar{u}_{ab} \bar{u}_{ba})$ and $
\Gamma_a = n_{imp} \frac{\sigma}{\pi N_a} = n_{imp} \pi N_b U_{ab} U_{ba} (1-\sigma)
$. The parameter $\sigma$ is used as an indicator of the strength of the impurity scattering. In the Born approximation $\sigma \to 0 $, while in the opposite unitary limit $\sigma \to 1$.

\end{document}